\documentstyle[epsfig]{aipproc}
\newcommand{\epsfigure}[2]{\epsfig{file=#1,width=#2}}

\begin{document}
\title{Saddles on the potential energy landscape of a Lennard-Jones liquid
\thanks{Invited talk presented at the Conference 
{\it Disordered and Complex Systems}, King's College London, July 2000.} 
}

\author{
Kurt Broderix$^*$, 
Kamal K. Bhattacharya$^*$, 
\underline{Andrea Cavagna}$^\dagger$, 
Annette Zippelius$^*$ and 
Irene Giardina$^{**}$
}

\address{
$^*$ Institut f\"ur Theoretische Physik, 
Universit\"at G\"ottingen, D-37073 G\"ottingen, Germany \\
$^\dagger$ 
Department of Physics and Astronomy, The University, Manchester,
M13 9PL, UK\\
$^{**}$ Service de Physique Th\`eorique, CEA Saclay, 
91191 Gif-sur-Yvette, France}

\maketitle

\vglue -1.0 truecm
\begin{abstract}
By means of molecular dynamics simulations, we study  
the stationary points of the potential energy in a Lennard-Jones liquid,
giving a purely geometric characterization of the energy landscape of 
the system. 
We find a linear relation between the degree of instability of
the stationary points and their potential energy, and we locate
the energy where the instability vanishes. This threshold 
energy marks the border between saddle-dominated and  minima-dominated 
regions of the energy landscape. 
The temperature where the potential energy of the Stillinger-Weber 
minima becomes equal to the threshold energy turns out to be very 
close to the mode-coupling transition temperature $T_c$. 
\end{abstract}
\vglue -0.6 truecm

The low temperature dynamics of supercooled liquids and glasses is
often put in relation with the geometric properties of the potential 
energy landscape of these systems. In particular, the presence of a
large number of inequivalent glassy minima has stimulated many studies
in the past \cite{cavagna-gold,cavagna-sw,cavagna-jo,cavagna-nat,cavagna-kk}.
More recently, the study of mean-field models of spin-glasses 
has strengthened the persuasion that the dynamical behaviour of 
glassy systems is deeply connected to the topology of the energy 
landscape \cite{cavagna-rev}. 
Moreover, it has been shown that spin-glass systems exhibiting  
one-step replica symmetry breaking (1RSB) have many dynamical properties in
common with fragile structural glasses \cite{cavagna-kb}, 
suggesting that 1RSB mean-field spin-glasses and fragile glasses may
have a similar energy landscape.

In this context, crucial questions are: How to characterize the energy
landscape of a glassy system? How to quantify the similarity of the
energy landscape of fragile glasses and 1RSB spin-glasses? 
Although utterly relevant, the structure of minima of 
the potential energy is not enough: even at low 
temperatures, when activation is the
only mechanism of diffusion in liquids, overcoming a barrier implies
crossing a {\it saddle} of the potential energy. Furthermore, at higher
temperatures the system spends more time around saddles than minima,
hence the structure of unstable stationary points 
is important for understanding the crossover from a
non-activated to an activated dynamics upon cooling
\cite{cavagna-madan,cavagna-bradipo}. 
The statistical properties of the stationary points of the potential
energy are of course independent of the temperature, as also should 
be a thorough description of the energy landscape itself.

Here we will focus on the purely geometric properties of the energy
landscape of a glassy system, by studying the statistical properties
of {\it all} the stationary points of its potential energy, be they
minima or saddles.  We classify them according to number of unstable
directions, (or {\it index} K), potential energy and smallest
eigenvalue of the Hessian matrix.  We show that in this way it is
possible to quantitatively compare the energy landscape of different
systems, pointing out similarities and differences. Moreover, we
discover a threshold energy below which it is highly unlikely to find
saddles and the relaxation requires activation. We thus establish a
connection between this threshold energy and the critical temperature
$T_c$ of mode-coupling theory (MCT) \cite{cavagna-mct}.  Finally, by
means of our results, we support some recent speculations on the role
of saddles in supercooled liquids \cite{cavagna-bradipo}.

The system under consideration is a binary mixture of Lennard--Jones (LJ)
particles \cite{cavagna-ka} (for details see \cite{cavagna-sad}).
Throughout this study we present results for systems with $N=60$ particles.
In order to explore the stationary points
of the potential energy, we use the following method: We 
equilibrate a configuration at a given temperature $T$ 
using a standard molecular dynamics (MD) simulation technique. 
To locate a saddle close to the equilibrium configuration 
we then perform a quench on a pseudo-potential energy
landscape $W(x)$ given by the modulus square of the force,
$W(x)=\vec\nabla U(x)\cdot\vec\nabla U(x)$, where $U(x)$ is the original
potential energy \cite{cavagna-sad,cavagna-ruocco}. 
All {\it absolute} minima of $W(x)$
are stationary points of $U(x)$, hence every saddle of $U(x)$ has a
well defined basin of attraction. The {\it local} minima of $W(x)$,
however, do not correspond to zeros of the real force. These points are
frequently sampled, but they can easily be distinguished from the 
absolute minima and are excluded from our analysis.
However, this means that the method does not associate to all 
the configurations a nearby saddle and therefore it cannot be used to 
build a natural dynamics of the relevant saddles, equivalent to the 
Stillinger-Weber (SW) one for minima \cite{cavagna-sw}. 
Furthermore, the relation between the initial MD equilibrium configuration at
temperature $T$ and the final stationary point found by this algorithm, is 
not straightforward. 
We prefer to perform here a purely geometric analysis
of the stationary points, independent of the way we have sampled 
them \cite{cavagna-foot}.

Given a stationary point, we compute its index density $k=K/(3N)$ and
its potential energy density $u=U/N$.  In Fig.1 we show
the results obtained by sampling saddles at two different values of
the temperature.  This plot clearly suggests that there is an
underlying curve $k(u)$ independent of the temperature, which encodes
a purely geometric feature of the landscape.  By sampling
stationary points at different values of $T$ we are simply exploring
different portions of the same geometric curve:
temperature acts as a light spot needed to unveil the underlying
function $k(u)$.
\begin{figure}[t] 
\begin{center}
\epsfigure{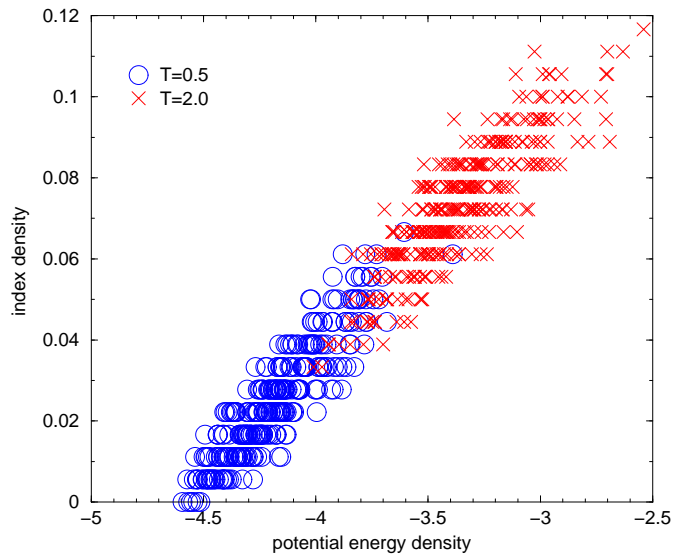}{3.5in}
\end{center}
\vspace{10pt}
\caption{Index density as a function of the potential energy density:
sampling of stationary points at two different temperatures, 
$T=0.5$ and $T=2.0$.} 
\end{figure}
In Fig.2 we show the average index density as a function of the
energy density. What is most striking of this plot is how well defined the 
function $k(u)$ is: due to its geometric nature there are no thermal 
fluctuations.
This curve shows that if we cut the potential energy landscape with a plane
of constant energy density $u=u_0$, the stationary points on this
plane (or within a narrow shell around this plane) will be dominated
by saddles with index density $k(u_0)$.
Furthermore, $k(u)$ is to a very good approximation a linear function
in the explored regime of $u$.  This implies that the curve
extrapolates to zero at a  well defined energy, which we call the {\it
threshold energy} $u_{th}$, in analogy with 1RSB spin-glasses.  
The linear interpolation of all the data and the linear
interpolation of the last four points give the same estimate for the
threshold, that is $u_{th}=-4.55$.
\begin{figure}[t] 
\begin{center}
\epsfigure{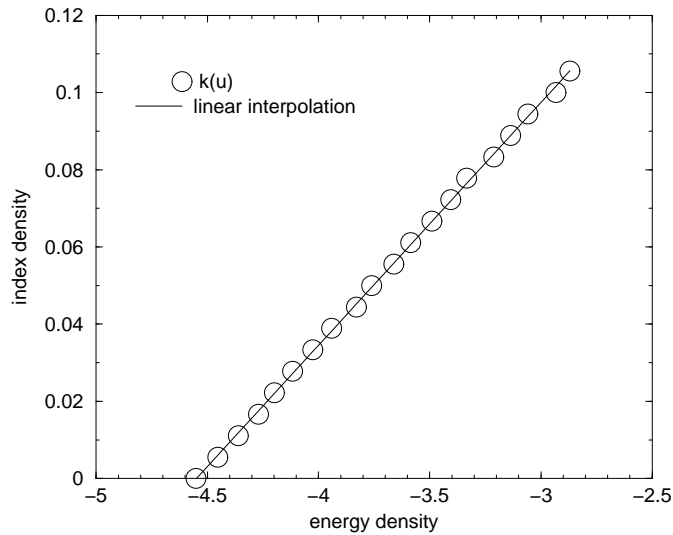}{3.5in}
\end{center}
\vspace{10pt}
\caption{Index density as a function of the potential energy density.
Average over all the data obtained by sampling at 
$T\in[0.3, 2.0]$. The full line is a linear fit of the data.}
\end{figure}

The threshold energy marks the border between the saddles-dominated 
portion of the energy landscape and the minima-dominated one.
A crucial point is that $u_{th}$ is {\it above} the energy of the 
deepest glassy minima found with the SW method
\cite{cavagna-sw}, that is $u_0=-4.65$: there is a finite energy
density interval, $u\in[u_0,u_{th}]$, where minima are entropically
dominant over saddles, as it happens in 1RSB spin-glasses 
\cite{cavagna-us3}. 
In those systems, however, $k(u)$ is $\it not$ 
a linear function and $k'(u)$ vanishes at the threshold. 
This difference may be related to the mean-field nature of 
1RSB models as opposed to real liquids. 
Indeed, in \cite{cavagna-bradipo} the slope of $k(u)$ has been 
connected to the energy barriers in the 
system by the relation $\Delta U\sim 1/k'(u)$: in the mean-field case we
expect barriers among minima to diverge, implying $k'(u_{th})=0$
(as found in 1RSB models), while this cannot be true in finite
dimensional systems.

To further compare the energy landscape of 
LJ and 1RSB systems, we consider in Fig.3 the lowest eigenvalue 
$\lambda_0$ of the Hessian in a stationary point, as a function of 
its potential energy density. 
As expected, $\lambda_0\to 0$ for $u\to u_{th}$, implying that the 
potential energy landscape at
the threshold has some flat directions, i.e. it is {\it marginal}
\cite{cavagna-rev}.
What is somewhat surprising is that $\lambda_0(u)$
is approximately a linear function of the energy, 
$\lambda_0\sim (u_{th} -u)$, exactly as in 1RSB spin-glasses 
\cite{cavagna-vira}. 
We conclude that both, the index density $k(u)$ and the smallest
eigenvalue $\lambda_0(u)$, provide a quantitative measure and allow
a direct comparison of the
properties of the energy landscape for two very different
systems, LJ liquids and 1RSB mean-field 
spin-glasses. 
\begin{figure}[t] 
\begin{center}
\epsfigure{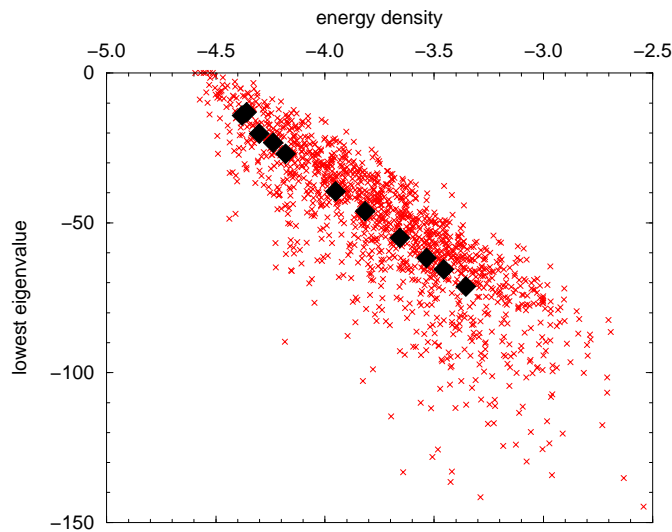}{3.5in}
\end{center}
\vspace{10pt}
\caption{Lowest eigenvalue of the Hessian as a function of the energy 
density. Full diamonds are the average of all the data obtained 
at the same values of $T$ as in Fig.2.}
\end{figure}

Our next task is to relate the threshold potential energy $u_{th}$  
to the dynamical behaviour of the supercooled liquid in the proximity
of the glass transition. 
In Fig.4 we plot the average energy density $u_{SW}(T)$ of the
SW minima as a function of the temperature of the initial 
MD trajectory \cite{cavagna-jo,cavagna-nat,cavagna-kk}, in
comparison to $\delta(T)\equiv\langle U(x)/N\rangle(T) - 3/2 T$, i.e.
the difference between the average potential energy density (from MD
simulations) and the vibrational energy in the harmonic approximation.
For a harmonic potential $\delta(T)$ is just the energy of the minimum
of the well. We find that $\delta(T)\sim u_{SW}(T)$ for $T\leq 1.2$
\cite{cavagna-sri}. This is the range of temperatures which is dominated by
the energy landscape and the timescales for the two processes of
relaxation - vibrations inside a minimum and hopping between different
minima - start to separate.  
Close to the glass transition (depending on the cooling
rate) the system falls out of equilibrium, as indicated by the 
saturation of both quantities, $u_{SW}(T)$ and $\delta(T)$. 
As we can see, the extrapolation of the equilibrium part of these
curves reaches the threshold energy at the MCT transition temperature $T_c$. 
The index density vanishes at the threshold, so that below this
energy minima are the entropically dominant stationary points.
Of course, there are minima also above the threshold, but they are
{\it not} dominant, while saddles are.
The MCT $T_c$ therefore corresponds to the temperature below which
minima visited by the dynamics become entropically dominant,
while saddles become statistically irrelevant.

\begin{figure}[t] 
\begin{center}
\epsfigure{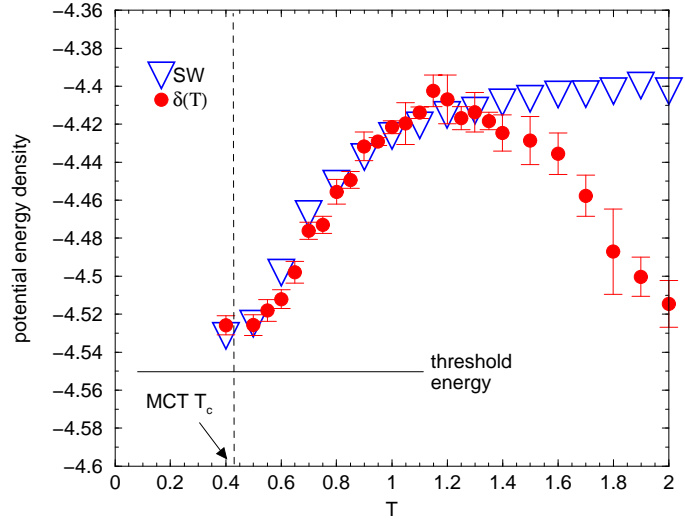}{3.5in}
\end{center}
\vspace{10pt}
\caption{Triangles represent the energy of the Stillinger-Weber minima 
$u_{sw}$ as a function of
the temperature of the initial MD trajectory. Circles represent the
quantity $\delta(T)$. The MCT transition is at $T_c\sim 0.44$.}
\end{figure}

To conclude, we consider the potential energy barriers between
different minima in our system.
According to \cite{cavagna-bradipo} we can obtain an estimate of the barriers 
from the slope of $k(u)$, as $\Delta U=1/[3 k^{\prime}(u_{th})]\sim 5.0$, 
which has the right order of magnitude (see, for example, the data of 
\cite{cavagna-sh}). 
Following the scenario of \cite{cavagna-bradipo}, we must locate the 
geometric crossover temperature $T_B$, below which saddles no longer 
contribute to diffusion: as we have seen above, this temperature must 
be identified in this system with $T_c$, giving $T_B\sim0.44$.
On the other hand, the other crossover temperature $T_A$, 
ruling activation, is fixed by the barrier size, $T_A\sim \Delta U\sim 5.0$. 
Hence, for this system we have $T_B < T_A$: at $T_B$ barriers are 
already quite large as compared to $T$, and following \cite{cavagna-bradipo}
this implies that the system is fragile. 
This result is consistent with the common 
classification of LJ liquids as fragile and therefore supports the 
description of fragile vs strong liquids behaviour given in 
\cite{cavagna-bradipo}.


\begin{references}
\bibitem{cavagna-foot} 
A quench on the pseudo-potential $W(x)$
is likely to decrease on average the real energy $U(x)$ as a 
consequence of the fact that the Hessian
has more positive than negative eigenvalues.
The increment of the energy during a
gradient descent step, $\vec dx=-\vec \nabla W\, dt$, 
is given by: $dU=\vec\nabla U\cdot \vec dx=
- \vec \nabla U \cdot \vec \nabla W\, dt=
-\vec \nabla U \cdot {\cal H} \cdot \vec \nabla U \, dt$, 
where ${\cal H}$ is the Hessian of $U$, which at low energies has 
only a small fraction of negative eigenvalues. Thus, if $\cal H$ and 
$\vec\nabla U$ are weakly correlated, we do expect the quadratic form
$\vec \nabla U \cdot {\cal H} \cdot \vec \nabla U$ to be positive on 
average, and therefore $dU$ to be negative.
\end{references}
\end{document}